\documentclass[preprint,12pt]{elsarticle}

\usepackage{amssymb}
\usepackage{mathrsfs}

\journal{Physics Letter B}

\begin{document}

\begin{frontmatter}



\title{Directional detection as a  strategy to discover  Galactic Dark Matter}


\author{J. Billard, F. Mayet, J.~F.~Mac\'\i as-P\'erez, D. Santos}

\address{Laboratoire de Physique Subatomique et de Cosmologie, Universit\'e Joseph Fourier Grenoble 1,
  CNRS/IN2P3, Institut Polytechnique de Grenoble, Grenoble, France}

\begin{abstract}
Directional detection of Galactic Dark Matter is a promising  search strategy for discriminating
WIMP events from  background. Technical progress on gaseous detectors and read-outs has permitted the design and construction of competitive experiments.
 However,  to take full advantage of this powerful detection method, one need to be able to 
  extract information from an observed recoil map to identify a WIMP signal. We present a comprehensive formalism, using a map-based likelihood method 
  allowing to recover the main incoming direction of the signal and its significance, thus  proving  its Galactic origin.  
  This is a blind analysis intended to be used on any directional data. 
Constraints are deduced in the ($\sigma_n, m_\chi$) plane and systematic  studies are presented in order to show that, using this analysis tool, unambiguous Dark Matter
 detection can be achieved on a large range of exposures and background levels. 
\end{abstract}


\end{frontmatter}


%
%
 
\newpage
\section{Introduction}
\label{sec:intro}

 A substantial body of astrophysical evidence supports the existence of non-baryonic Dark Matter :  locally,  
from the rotation curves of spiral galaxies \cite{rubin}, on cluster scale from 
 the observation of galaxy cluster collisions \cite{clowe} and on the largest scale from  cosmological  
 observations \cite{wmap}. Candidates for this class of   to-be-discovered particles (WIMP: Weakly Interacting Massive Particles) naturally arise from 
extensions of the standard model of particle physics, such as supersymmetry~\cite{susydm} or extra-dimensions~\cite{kkdm}.\\ 
Like most spiral galaxies, the Milky Way is supposed to be immersed in a halo of WIMPs which outweighs the luminous component by at 
least an order of magnitude. 
Tremendous experimental efforts on a host of techniques have been made in the field of direct search of 
WIMP Galactic Dark Matter. 
With really small expected event rates (${\cal O} \rm (10^{-5}-1) \ day^{-1}kg^{-1}$), 
the challenge is to  distinguish  a genuine WIMP signal from  background events.
Two main Dark Matter search strategies may be identified : the detector may be designed to reach an  
extremely  high level of background rejection \cite{cdms,revue} and/or to provide an unambiguous positive WIMP signal.
The second strategy can be applied by searching for a correlation of the WIMP signal either with the motion of the Earth around the Sun, observed as an annual 
modulation of the number of events \cite{freese}, or with the direction of the Solar motion around the Galactic center \cite{spergel},
which happens to be roughly in the direction of the Cygnus constellation. 
The latter is generally referred to as directional 
 detection of Dark Matter and several project of detectors are being developed for 
this goal \cite{Drift,newage,MIMAC,mit,white}.\\ 
Directional detection of Dark Matter has first been proposed in \cite{spergel}, highlighting the fact that a strong forward/backward 
asymmetry is expected in the case of an isothermal spherical Galactic halo, which could, in principle, be observed   even with a poor angular
resolution. 
Since then, several phenomenological studies have been performed \cite{copi1,copi2,copi3,morgan1,morgan2,green1,green2,
host,vergados,gondolo}.  
Using an unbinned likelihood  method \cite{copi1,copi2,copi3} or non-parametric statistical tests on unbinned data 
\cite{morgan1,morgan2,green1,green2}, 
it has been shown that a few number of events  ${\cal O} (10)$ is required to reject the isotropy, which is characteristic of the background event distribution.
However, these methods do not allow to identify and characterize a WIMP signal and hence do not take 
full advantage of data from upcoming directional detectors \cite{white}.

In this Letter,  we use a map-based likelihood analysis in order to extract from an 
observed recoil map the main incoming direction of the events and its significance. In this way, the Galactic 
origin of the signal can be proved by showing its correlation with the direction of 
the Solar motion with respect to the Dark Matter halo.  Hence, the goal of this new method is not to reject the background hypothesis, but rather to identify a genuine WIMP signal. This method is tested on 
realistic simulated cases defined by a low number of WIMP events ${\cal O} (10)$ hidden  by a non-negligible background   and 
measured with a rather low angular resolution $\sim 15^\circ$ (FWHM) \cite{santos}.  
This blind analysis is intended to be applied to directional data of any detector.\\
The outline of this Letter is the following : after introducing the directional detection framework and phenomenology, the  map-based likelihood
analysis is presented and applied to a realistic simulated recoil map. Main results are presented and  constraints are 
deduced in the  $(\sigma_n, m_\chi)$  plane. 
Eventually, systematic  studies are presented in order to show that this analysis tool gives satisfactory results on a large 
range of exposures and background levels.

\section{Directional detection framework}
\label{sec:theory}

\subsection{Directional detectors}
\label{sec:detect}
 
Several directional detectors are being developed and/or operated :  DRIFT \cite{Drift}, NEWAGE~\cite{newage},
MIMAC~\cite{MIMAC}, DM-TPC~\cite{mit}. A detailed overview of the status of experimental efforts devoted to directional
Dark Matter detection is presented in \cite{white}. Directional detection of Dark Matter requires track reconstruction  of recoiling nuclei down 
to a few keV. This can be achieved with low pressure gaseous detectors \cite{sciolla} and several gases have been suggested : 
$\rm  CF_4,^{3}He+C_4H_{10}$ or  $\rm CS_2$. 
Both the energy and the track of the recoiling nucleus need to be measured precisely. Ideally, recoiling tracks should be 3D 
reconstructed as the required exposure  is decreased by an order of magnitude between  2D read-out and 3D read-out \cite{green1}. 
Sense recognition of the recoil track ({\it head-tail}) is also a key issue for
directional detection \cite{headtail1,headtail2}.

\subsection{Theoretical framework}
As a first application of the method, we focus on the simplest model for the Milky Way halo :  the isotropic isothermal sphere, 
in which the WIMP velocity follows
 a Maxwellian distribution defined in the laboratory rest frame as 
\begin{equation}
f(\vec{v}) = \frac{1}{(2\pi\sigma^2_v)^{3/2}}\exp\left (-\frac{(\vec{v} + \vec{v}_{\odot})^2}{2\sigma_v^2}\right )
\end{equation}
with a dispersion $\sigma_v = v_0/\sqrt{2}$ where $v_0 = 220 $ km.s$^{-1}$
  is the circular speed at Solar radius. As an illustration of the method, we consider a detector velocity equal to the 
  tangential component of the Sun motion around the Galactic center
$v_{\odot} = 220 \pm 20$ km.s$^{-1}$, neglecting the Sun peculiar velocity and the Earth orbital velocity about the Sun\footnote{Obviously, when analysing real data, these two components of the detector velocity have to be considered in order
 to have an accurate analysis.}. As a matter of fact, 
their contribution  to detector velocity 
is less than the uncertainty on $v_{\odot}$  \cite{morgan1}.  
Using the Galactic coordinates  $(\ell, b)$, the WIMP  velocity is written in the Galactic rest frame as~:
$$\vec{v} = v\ (\cos \ell \cos b \ \hat{x}  + \sin \ell \cos b \ \hat{y} + \sin b \ \hat{z})$$
where $\hat{x}$ points towards the Galactic center, $\hat{y}$ in the direction of the Solar motion and  $\hat{z}$ towards the Galactic north pole.  
The recoil distribution is then computed by generating random incident WIMP velocities from $vf(\vec{v})$ and assuming 
an isotropic elastic scattering in the center of mass frame. Then, the recoil energy ($E_R$) in the laboratory rest frame is given by
\begin{equation}
E_R = \frac{2v^2m^2_{\chi}m_N}{(m_{\chi} + m_N)^2}\cos^2 \theta_R
\label{RecoilEnergy}
\end{equation}
with  $m_{\chi}$ the WIMP mass, $m_N$ the mass of the target and $\theta_R$ the recoil angle in the laboratory frame. Within this framework, the WIMP signal is
expected to come from the Solar direction ($\ell_\odot = 90^\circ,  b_\odot =  0^\circ$) to which points the $\hat{y}$ 
axis\footnote{However, when considering the peculiar velocity of the
Sun, which is in the Galactic coordinate $\vec{v}_{\odot_p} = (~10.0,~7.3,~5.2)$ km.s$^{-1}$, 
the Solar motion direction is ($\ell_\odot = 87.5^\circ,  b_\odot =  1.3^\circ$)}. This
 happens to be roughly in the direction of the Cygnus constellation.\\
In the following of the Letter, we will consider a form factor $F(E_R)$ taken to be equal to one. Indeed, using the Born approximation, in the case of a spin-dependent interaction, the
form factor is given by the Fourier transform of a thin shell \cite{lewin} leading to: $F^2({\rm 5~keV}) = 0.99$ and $F^2({\rm 50~keV}) =
0.9$ in the case of a $\rm ^{19}F$ target, justifying our approximation. Note that in the case of heavier targets, this approximation would be no longer valid.

\subsection{Recoil maps}
Recoil distributions are presented in Galactic coordinate maps, 
using the HealPix~\cite{healpix} tool in order to respect the spherical topology and to have an overall sky map with equal area bins.  
Figure \ref{fig:DistribRecul} presents on the upper panel the theoretical WIMP flux and on the middle one, the theoretical 
recoil distribution evaluated with $10^8$ WIMP-induced events, in the case of a $^{19}$F target, a  100 GeV.c$^{-2}$ WIMP 
 and considering recoil energies in the range  5 keV $\leq E_R \leq$ 50 keV. The lower bound of the energy range is due to the threshold ionization energy taking into account the
 quenching factor. As most of the WIMP events are concentrated at low recoil energy, an upper bound is
 chosen to limit the background contamination of the data. Indeed, following this framework, 70\% of the recoils are between 5 keV and 50 keV and only 10\% above 50 keV. 
 Thus, increasing the upper bound would lead to a potentially weaker signal to noise ratio. 
   It can be noticed that after  scattering, the WIMP-induced recoil distribution, although obviously broadened, 
 is still pointing in the Solar motion direction ($\ell_\odot,  b_\odot$). Thus, the expected WIMP signal is clearly 
 anisotropic and directional detection should be able to 
 distinguish a genuine WIMP signal from an isotropic background.

\subsection{WIMP signal characteristics}
\label{sec:WIMPSignal}
\begin{figure}[p]
\begin{center}
\includegraphics[scale=0.45,angle=270]{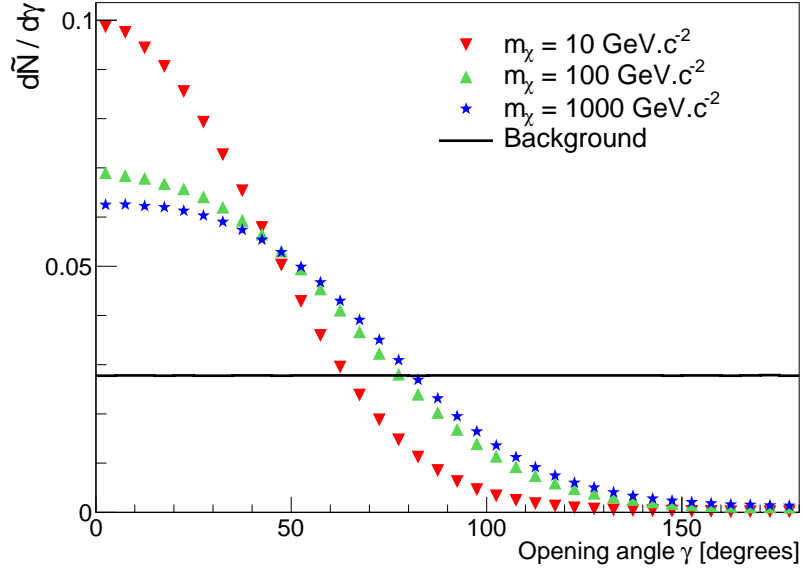}\\
 \end{center}
\vspace*{5mm} 
 \begin{center}
\includegraphics[scale=0.45,angle=270]{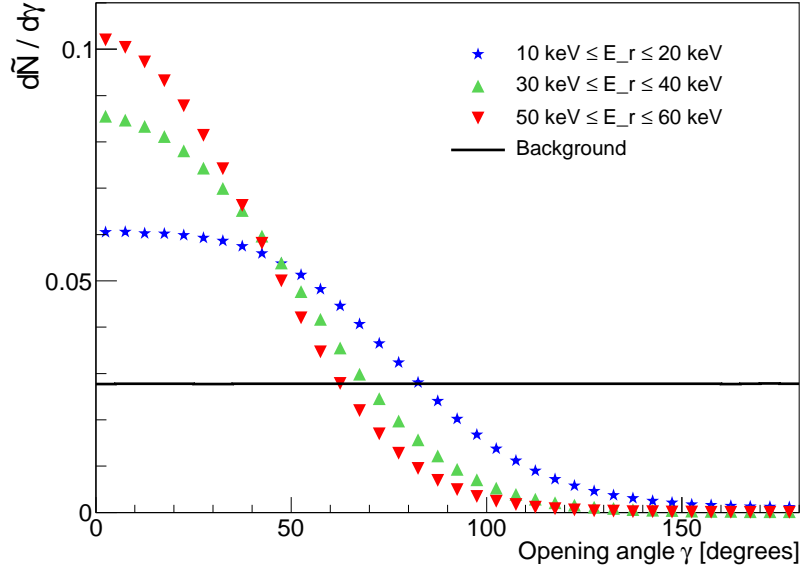}
\caption{Normalized angular spectra of the WIMP signal,   i.e.   fraction of events per solid angle as a function of the opening angle 
$\gamma$, defined as the angle between the $\hat{y}$ axis and the recoil direction. All plots are in the case of a $^{19}$F target.
Upper panel presents the normalized angular spectra for three different values of $m_{\chi}$ and  a recoil energy range 5 keV $\leq E_R \leq$ 50 keV. 
Lower panel presents the normalized angular spectra  for three different recoil energy ranges and  with $m_{\chi}$ = 100 GeV.c$^{-2}$.
Black solid line presents the isotropic background spectrum.}  
\label{fig:distrib.angul}
\end{center}
\end{figure}

In order to evaluate the evolution of the angular distribution of WIMP events as a 
function of the WIMP mass ($m_{\chi}$) and the  recoil energy range ($E_R$), 
we first look at the 1D angular spectrum, as it is a  convenient representation of the recoil map. 
The latter is defined as the normalized  fraction of events per solid angle as a function of the opening angle 
$\gamma$,  corresponding to the angle between the $\hat{y}$ axis and the recoil direction. The result is twofold~:
\begin{itemize}
\item The upper panel of figure \ref{fig:distrib.angul} presents the 1D angular spectrum of WIMP events for three 
different values of $m_{\chi}$: 10, 100, 1000 GeV.c$^{-2}$ in the case of a $^{19}$F target for a recoil energy 
in the range 5 keV $\leq E_R \leq$ 50 keV.  We can notice that the WIMP event angular distribution clearly depends on the WIMP mass.
 Lighter is the WIMP, stronger is the angular anisotropy. Indeed, due to the finite energy range and the fact that 
 low   WIMP mass induce an energy distribution  shifted to low energy,  events above threshold are the one with the most directional feature (eq. (\ref{RecoilEnergy})). 
 The directionality evolves quickly at low masses and very slowly for masses heavier than a hundreds of GeV.c$^{-2}$. 
 Even if the WIMP is heavy ($\sim 1$ TeV.c$^{-2}$), the signal is 
 still directional and then  different enough from the background to be identified.  
Hence, directional detection presents the potential to   distinguish a
WIMP signal from background, for any WIMP mass. 
\item The lower panel of figure \ref{fig:distrib.angul} presents, in the case of a $^{19}$F target, three different 
1D angular spectra corresponding to three 
different recoil energy ranges: 10 keV $\leq E_R \leq$ 20 keV, 30 keV $\leq E_R \leq$ 40 keV and
50 keV $\leq E_R \leq$ 60 keV. We can notice that the whole angular distribution also depends on the recoil 
energy range and that greater is the recoil energy, stronger is the
 angular signature, see eq.  (\ref{RecoilEnergy}).
\end{itemize}

The recoil map does also depend on the target mass $m_N$ and as several directional detectors are being developed with different 
targets, we study the influence of
these targets. Although their detection characteristics may be different (e.g. track length, drift velocity and straggling\footnote{This is the angular deviation due to
scatterings of the recoiling nucleus with other nuclei from the gas.}), for low mass targets ($\rm H$, $\rm ^{19}F$, $\rm ^3He$) and at sufficiently low recoil energy when the form
factor can be approximated to unity,  equivalent directional signal can be found by  
adjusting the energy range for each target.  
 Indeed, the directionality of the signal is encoded only  
in the $\cos^2 \theta_R$ term of eq. (\ref{RecoilEnergy}), then  the   angular distribution  for a target  of  mass $m_{N_1}$ at a recoil energy $E_{R_1}$ is
equivalent to the one of a $m_{N_2}$ target at $E_{R_2}$, using : 
\begin{equation}
 E_{R_2} =  E_{R_1}\frac{m_{N_2}}{m_{N_1}}  \left ( \frac{m_{\chi}+m_{N_1}}{m_{\chi}+m_{N_2}} \right )^2
\end{equation} 
Of course, this equivalence relation is no more valid for heavier targets when the form factor is far from unity.
Hereafter, we consider
  a $^{19}$F target with recoil energy in the range $5 \leq E_R \leq 50$
 keV  and $m_{\chi} = 100$ GeV.c$^{-2}$ (fig. \ref{fig:DistribRecul}), but same WIMP-induced recoil distribution would be obtained  for 
 $\rm ^{3}He$ target with  $1 \leq E_R \leq 10$ keV.

\section{Map-based likelihood analysis}
\label{sec:like}

\subsection{A realistic simulated measurement}
Lower panel of figure \ref{fig:DistribRecul} 
presents a typical recoil distribution observed by a directional detector : $100$ WIMP-induced events and 
$100$ background events generated isotropically.  These   events  are meant to be after data rejection based e.g. on track length and energy selection \cite{MIMAC}. 
For an elastic axial cross-section on nucleon $\rm \sigma_{n} = 1.5 \times 10^{-3} \ pb$ and a $\rm 100 \ GeV.c^{-2}$ WIMP mass, this corresponds to 
an exposure of $\rm \sim 7\times 10^3  \ kg.day$ in  $\rm ^{3}He$ and $\rm \sim 1.6 \times 10^3 \ kg.day$  in CF$_4$, 
on their equivalent energy ranges discussed in sec.~\ref{sec:WIMPSignal}.
  Low resolution maps are used in this case ($N_{\rm pixels} = 768$) which is sufficient  for a rather low 
  angular resolution, $\sim 15^\circ$ (FWHM), expected for this type of detector and   justified  for instance by the  
straggling of the recoiling nucleus \cite{santos}. 
3D read-out and sense recognition are considered for this study.\\

\subsection{Likelihood definition}
\begin{figure}[p]
\begin{center}
\includegraphics[scale=0.35,angle=90]{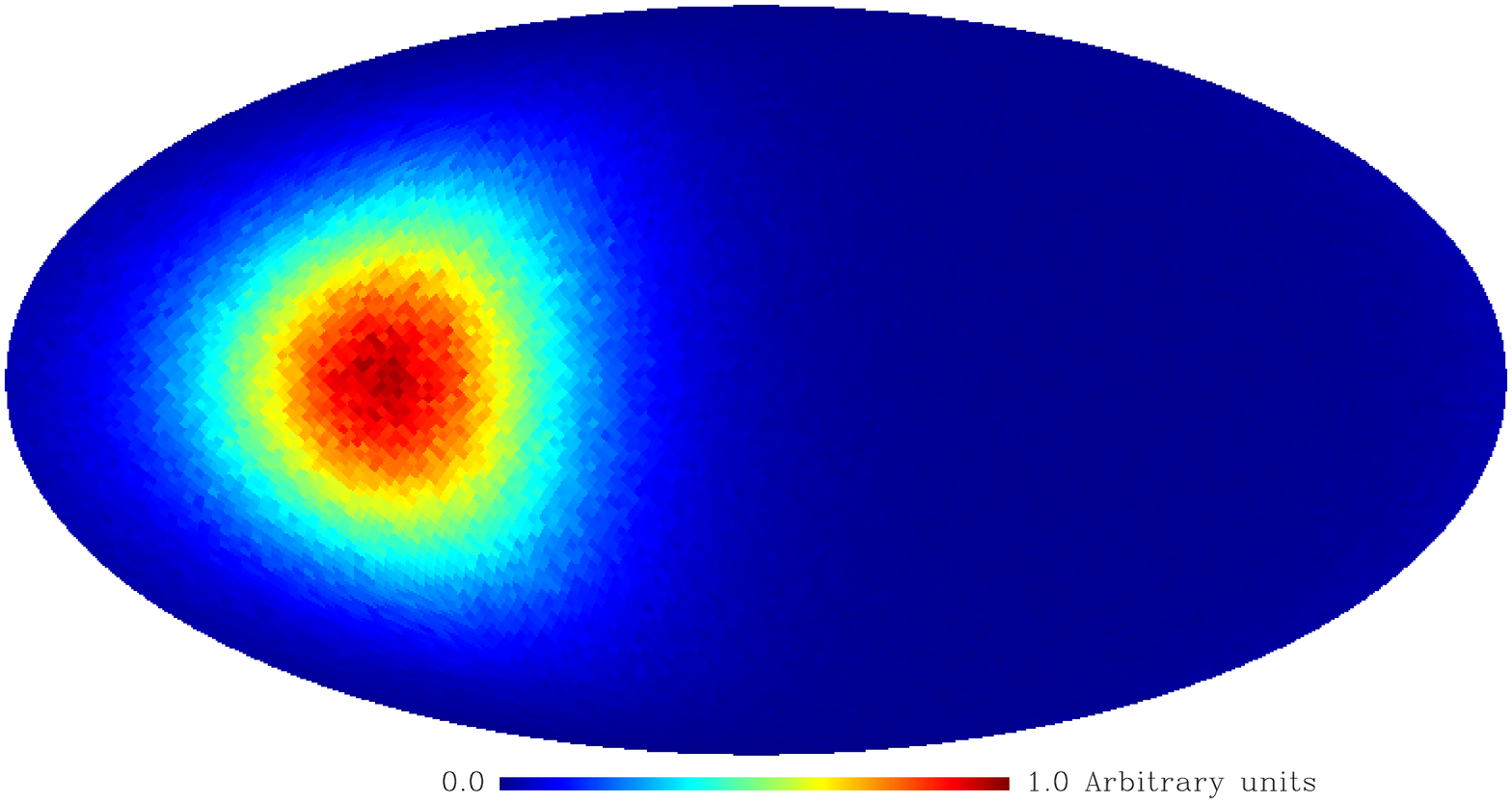}
\includegraphics[scale=0.35,angle=90]{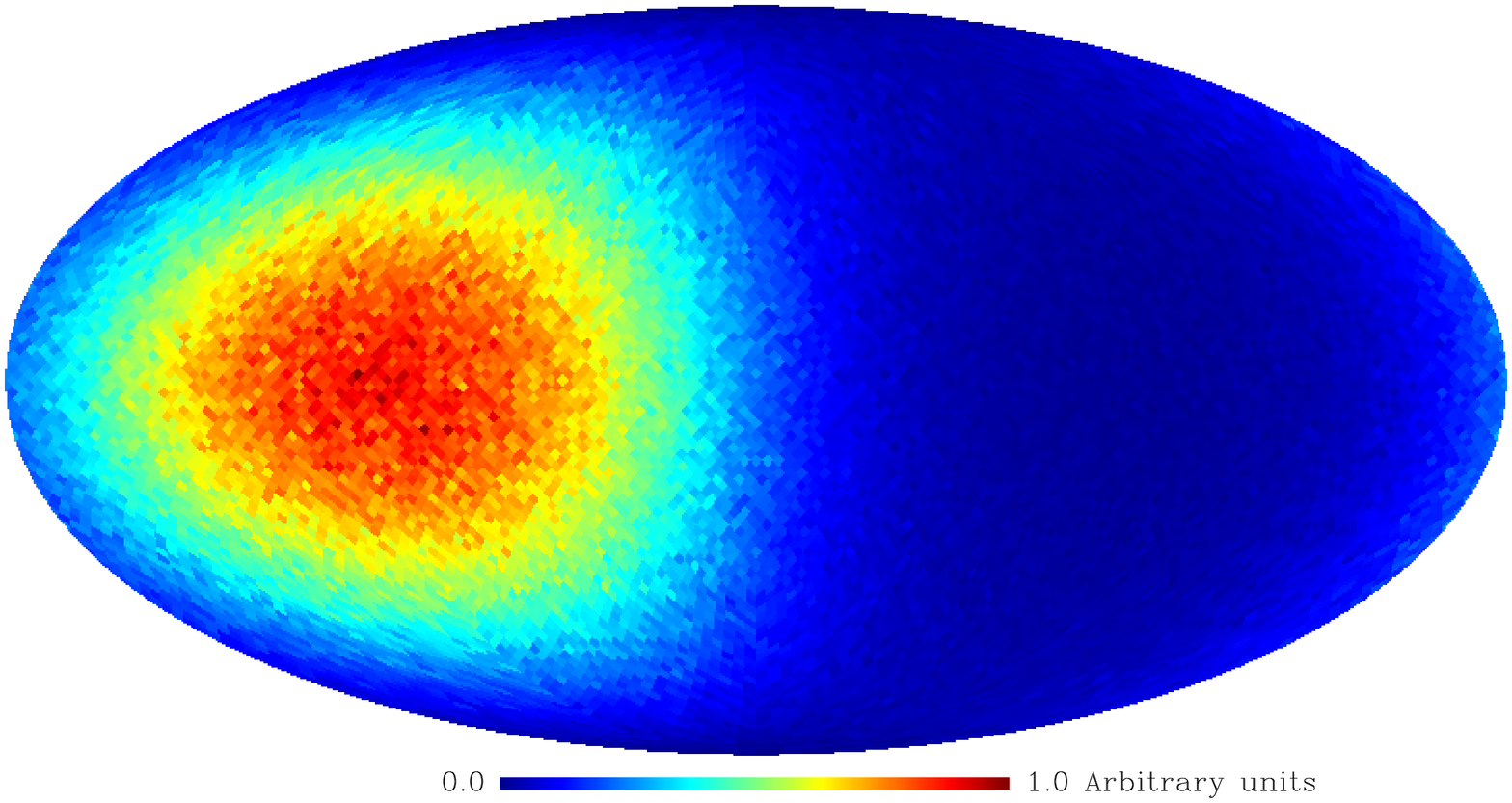}
\includegraphics[scale=0.35,angle=90]{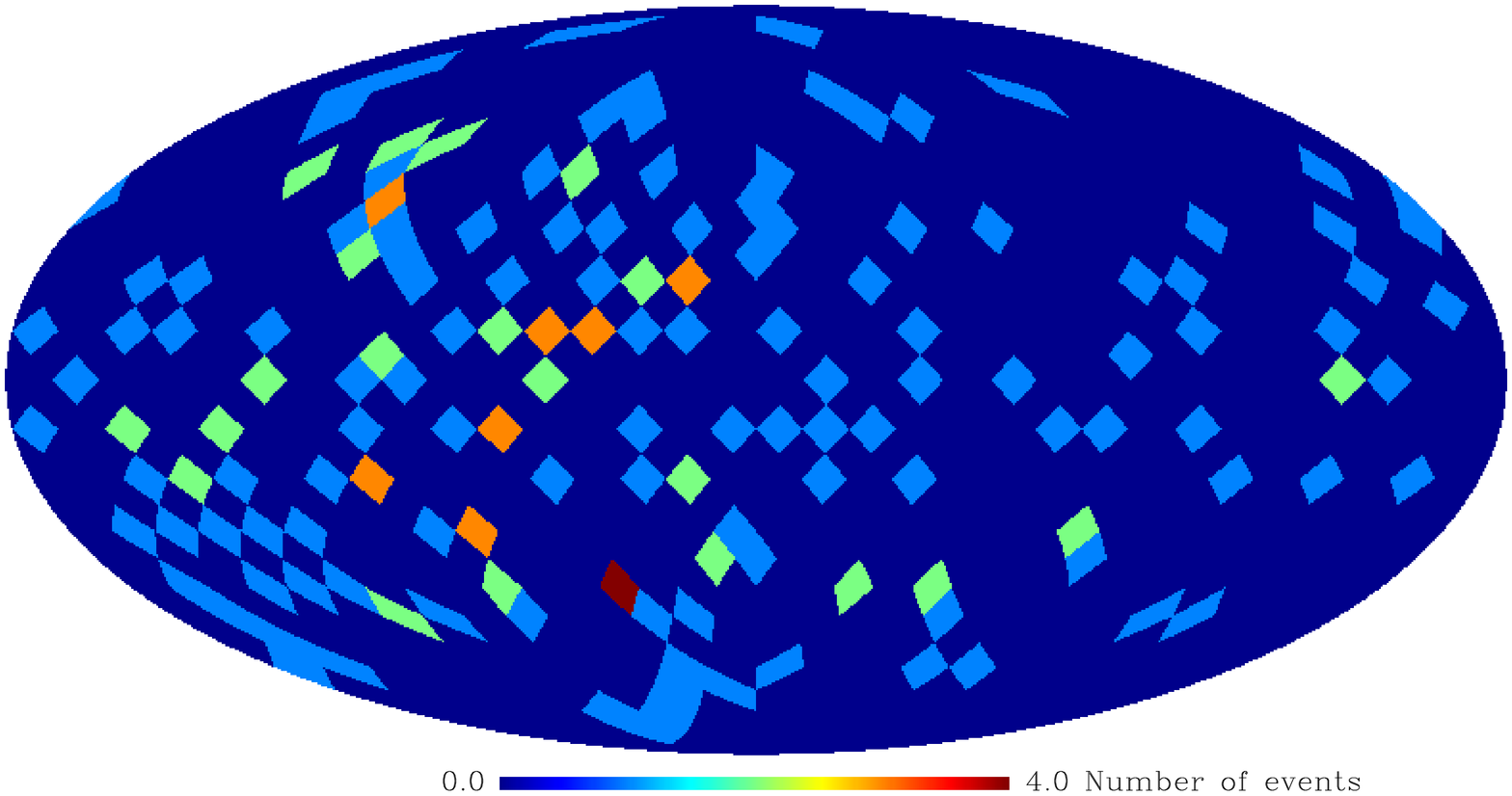}
\caption{From top to bottom : WIMP flux  in the case of an isothermal spherical halo,   WIMP-induced recoil distribution 
 and a typical simulated measurement :  100 WIMP-induced recoils and 100 background
events with a low angular resolution. Recoils maps are produced for a $^{19}$F target, a  100 GeV.c$^{-2}$ WIMP 
 and considering recoil energies in the range  5 keV $\leq E_R \leq$ 50 keV. Maps are Mollweide equal area projections.}  
\label{fig:DistribRecul}
\end{center}
\end{figure}

At first sight, it seems difficult to conclude from the recoil map of fig.~\ref{fig:DistribRecul} that it does contain 
a fraction of WIMP events pointing towards the direction of the Solar motion. 
A likelihood analysis is developed in order to retrieve from a recoil map : the main direction of the incoming events in 
Galactic coordinates ($\ell, b$) and the number of WIMP events contained in the map. The likelihood value is estimated using a binned map 
of the overall sky with  Poissonian statistics,  as follows :
 \begin{equation}
 \mathscr{L}(m_\chi,\lambda, \ell,b) = \prod_{i=1}^{N_{\rm pixels}} P( [(1-\lambda) B_i + \lambda S_i(m_\chi ;\ell,b) ]|M_i)
 \end{equation}
where $B$ is the  background spatial distribution 
taken as isotropic, $S$ is the WIMP-induced recoil distribution and $M$ is the measurement. 
This is a four parameter likelihood analysis with $m_\chi$, 
 $\lambda = S/(B+S)$ the  WIMP fraction (related to the  background 
rejection power of the detector) and the coordinates ($\ell$, $b$) referring to the maximum of the 
WIMP event angular distribution.
Hence, $S(m_\chi;\ell,b)$ corresponds to a rotation of the $S(m_\chi)$ distribution 
by the angles ($\ell' = \ell - \ell_\odot$, $b' = b - b_\odot$).

A scan of the four parameters with flat priors, allows to evaluate the likelihood between the measurement 
(fig.~\ref{fig:DistribRecul} bottom) and the theoretical distribution made of a superposition of 
an isotropic background and a pure WIMP signal (fig. \ref{fig:DistribRecul} middle). By scanning on $\ell$ and $b$ values, we ensure 
that there is no prior on the direction of the center of the WIMP-induced recoil distribution. In order to respect the spherical topology, a careful rotation of the $S$ distribution
 on the whole sphere must be done as follows.
 Given $\overrightarrow{V}_i$ the vector pointing on a 
bin $S_i$, the following rotation is considered : 
$$\overrightarrow{V}^\prime_i = R_{\vec{u}}(b')R_{\hat{z}}(\ell')\overrightarrow{V}_i$$
 with $\vec{u} = R_{\hat{z}}(\ell') \ \hat{x} = u_x \ \hat{x} + u_y \ \hat{y}$ and $R_{\vec{u}}(b')$  the  Rodrigues 
 rotation matrix around an arbitrary 
vector  $\vec{u}$ given by : 

\begin{equation}
\left (
 	\begin{array}{ccc}
	\cos b' + u^2_x (1-\cos b') & u_x u_y (1-\cos b') & u_y \sin b'  \\	 
	u_x  u_y(1-\cos b') & \cos b' + u^2_y  (1-\cos b') & -u_x  \sin b' \\
	-u_y  \sin b' & u_x  \sin b' & \cos b' 
	\end{array} \right )
\end{equation}
where the angles are defined as follows : $\ell' = \ell - \ell_\odot$, $b' = b - b_\odot$.\\
 
The events contained in the observed recoil map can be either   WIMP-induced recoils or  background events. A realistic recoil map from
upcoming directional detectors cannot be background free. With this method, both components (background and signal) are taken into account and  
no assumption on the origin of each event is needed. Indeed, the observed map  is considered as a superposition of  
the background and WIMP signal distributions, and the likelihood method allows to recover $\lambda$, the signal to noise ratio. 
The advantage
is twofold :
\begin{itemize} 
\item First, background-induced bias is avoided. This would not be the case with a method trying to 
evaluate a  likelihood on a map containing a fairly large number of background events considering only a pure WIMP reference distribution.  
\item Second, the value of $\lambda$ allows to access  the number of genuine WIMP events
and consequently to constrain the WIMP-nucleon cross-section as presented in sec. \ref{sec:cross}.
\end{itemize}

It is worth noticing that the likelihood is performed on the whole angular distribution in order to maximize the information contained 
in the observed recoil map leading to   more restrictive constraints on the four parameters. For instance, 
working on the forward/backward asymmetry is not sufficient to constrain the
($\ell, b$) parameters which are indeed the ultimate proof of the Galactic origin of the signal.

\subsection{Results from a realistic recoil map}
\begin{figure}[p]
\begin{center}
\includegraphics[scale=0.55,angle=270]{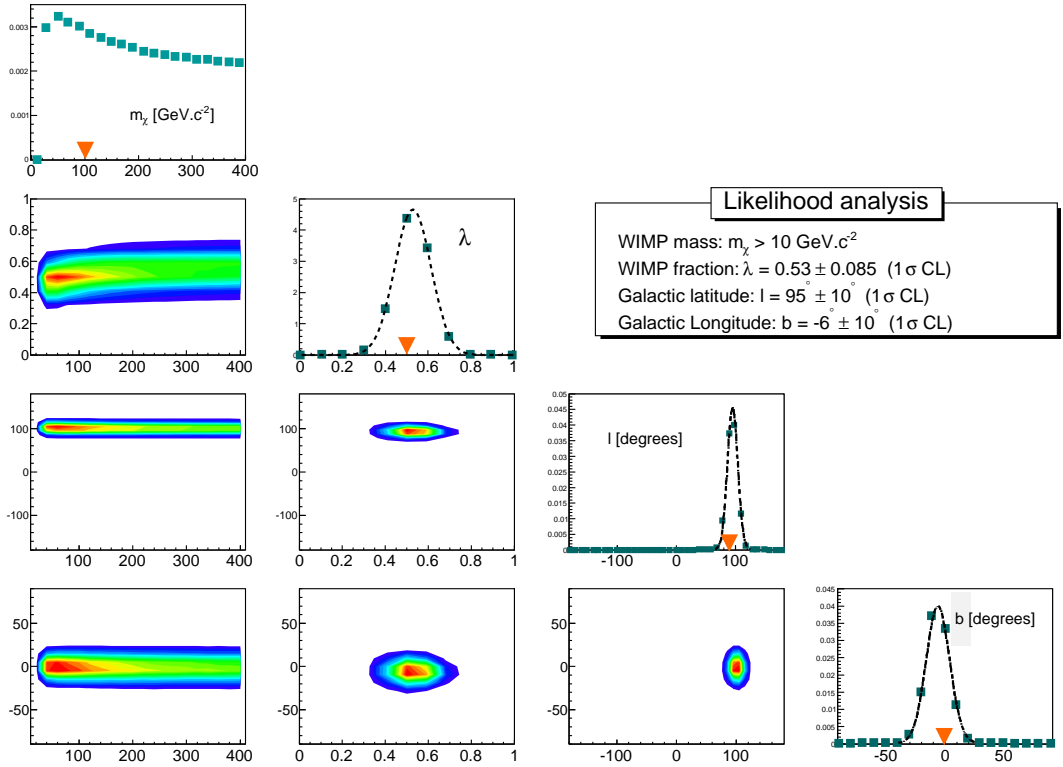}
\caption{Marginalised distribution (diagonal) and 2D distribution (off-diagonal) plots  of the four 
parameters $m_\chi , \lambda, \ell, b$   from the analysis of  
the simulated map (fig. \ref{fig:DistribRecul} lower panel).  Blue squares represent the calculated values of $\mathscr{L}$ 
and orange triangles refer to the input value of each parameter:   
$m_{\chi} = 100 \ {\rm GeV.c}^{-2},  \lambda = 0.5, \ell = 90^\circ, b = 0^\circ$.}  
\label{fig:LikelihoodTotal1}

\end{center}
\end{figure}

The four parameter likelihood analysis has been computed on 
the simulated map (fig. \ref{fig:DistribRecul} bottom) and the full result is presented on figure \ref{fig:LikelihoodTotal1}. 
Marginalised distribution (diagonal) and 2D distribution (off-diagonal) plots  of the four 
parameters $m_\chi , \lambda, \ell, b$ are presented.
 The conclusion of the analysis is twofold:
 \begin{itemize}
\item Firstly, in the lower half part of   figure \ref{fig:LikelihoodTotal1} are presented the marginalised distributions of the parameters $\ell$, $b$ and the 2D
representation of $\mathscr{L}$ in the ($\ell$, $b$) plane.
 These two parameters are well constrained  and the first result of this  map-based likelihood method is that  
the recovered main recoil direction is pointing towards 
($ \ell = 95^{\circ} \pm 10^{\circ}, b = -6^{\circ} \pm 10^{\circ}$) at 
$68 \  \%$ CL, corresponding to a non-ambiguous detection of particles from the Galactic halo. This is indeed the discovery  
proof of this detection strategy.
\item Secondly, on the upper half part of figure \ref{fig:LikelihoodTotal1} are presented the marginalised distributions of the parameters $m_\chi$, $\lambda$ and the 2D
representation of $\mathscr{L}$ in the ($m_\chi$, $\lambda$) plane. 
The method allows to constrain $\lambda$ the WIMP fraction in the observed recoil map but not the WIMP mass. Constraining the WIMP mass 
is not the main point of this analysis
tool. In fact, $m_{\chi}$ is set as a free parameter in order to show that the analysis is particle physics model independent. 
Indeed, 
the unknown value of $m_\chi$ does
not affect the values of $\lambda$, $\ell$ and $b$ due to the absence of correlations. Then, we can see on figure \ref{fig:LikelihoodTotal1}
 that only $m_\chi \leq 10$ GeV.c$^{-2}$ is excluded.
 We can estimate the number of WIMP events as $N_{\rm wimp} = \lambda \times N_{tot}$ where 
$N_{tot} = S + B$ follows a Poissonian statistic, and $\Delta N_{\rm wimp}$ is given by 
$\Delta N_{\rm wimp} \approx \Delta \lambda \times N_{tot}$.
\end{itemize}
 As a conclusion of this likelihood analysis, we can see that this method allows to determine from a realistic simulated   recoil map (fig. \ref{fig:DistribRecul}
  bottom) that it does contain a signal pointing towards the Cygnus constellation within 10$^\circ$, with $N_{\rm wimp}=106 \pm 17 \ (68 \% {\rm CL})$, corresponding to a high
   significance detection of Galactic Dark Matter.


\subsection{Constraining the elastic scattering cross-section}
\label{sec:cross}
\begin{figure}[p]
\begin{center}
\includegraphics[scale=0.78]{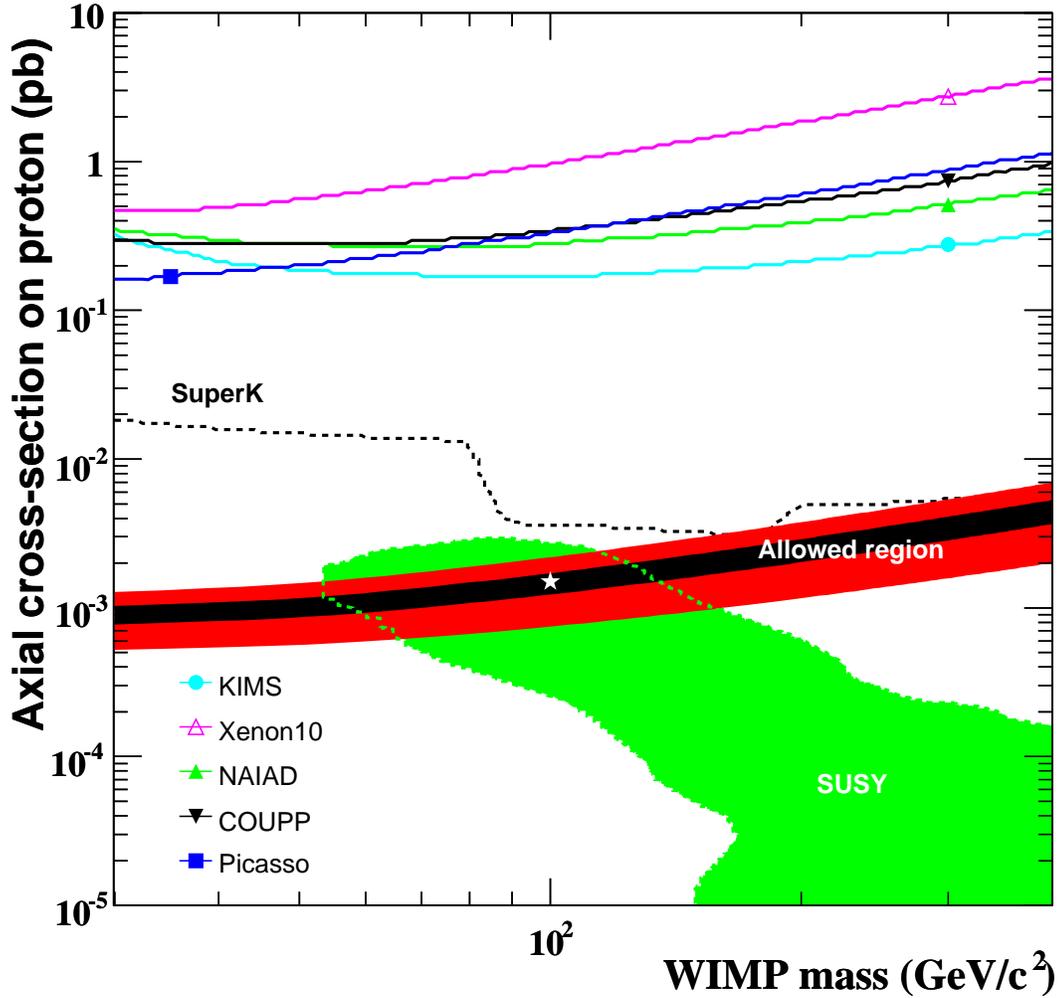}
\caption{Spin dependent cross-section on proton (pb) as a function of the WIMP mass ($\rm GeV/c^2$). Results are presented
in the case of pure-proton approximation \cite{tovey}. 
The theoretical region, obtained within the framework of the constrained minimal 
 supersymmetric model, is taken from \cite{superbayes}. Constraints from collider data and relic abundance are accounted for.
Exclusion limits from direct detection searches are presented :   COUPP~\cite{coupp},  
KIMS~\cite{kims}, NAIAD~\cite{naiad}, Picasso~\cite{picasso} and  Xenon10~\cite{xenon10}. 
Exclusion limit obtained with the Super-K neutrino telescope \cite{superk} is also displayed.
The allowed regions obtained with the example map shown are presented with shaded area,  with $1\sigma$ and $3\sigma$  
CL contours.  Input value for the simulation is shown with a star.}
\label{fig:discovery}
\end{center}
\end{figure}

A constraint in the $(\sigma_n, m_\chi)$ plane is then  deduced from  
the marginalised $\mathscr{L}(\lambda)$ distribution evaluated for each WIMP mass above $10$ GeV.c$^{-2}$. 
Using  the  standard expression of the event rate with a form factor $F^2(E_R)$ taken equal to one and the local 
halo density $\rho_0 = 0.3$ GeV.c$^{-2}$.cm$^{-3}$,
the $1\sigma$ and $3\sigma$  CL contours are calculated. 
Figure \ref{fig:discovery} presents the spin dependent  cross-section on proton (pb) as a function of the 
WIMP mass ($\rm GeV/c^2$). Results are presented
in the case of a   pure-proton approximation \cite{tovey} and the  proton spin content of $^{19}$F has been chosen as $<S_p>=0.441$ 
\cite{pacheco}. The $1\sigma$ and $3\sigma$ contours deduced from the analysis of the simulated recoil map are presented as 
shaded areas.  It should be highlighted that these contours represent the allowed regions, as directional detection aims at identifying WIMP signal 
rather that rejecting the background.  Exclusion limits from direct detection searches are presented :   COUPP~\cite{coupp},  
KIMS~\cite{kims}, NAIAD~\cite{naiad}, Picasso~\cite{picasso} and  Xenon10~\cite{xenon10}. 
Exclusion limit obtained with the Super-K neutrino telescope \cite{superk} is also displayed. 
 The theoretical region, 
obtained within the framework of the constrained minimal 
supersymmetric model,  is taken from \cite{superbayes}. Constraints from collider data and relic abundance $\Omega_\chi h^2$,
 as measured with WMAP 5-year data \cite{wmap5}, are accounted for. The input value for the simulation is presented as a star. 
\\ 
Such a  result could be obtained, with a background rate of 
$\sim 0.07$ kg$^{-1}$day$^{-1}$ and a  10 kg $\rm CF_4$ detector 
during  $\sim 5$ months, noticing that the detector should allow 3D 
track reconstruction, with sense recognition down to 5 keV. 

\section{Discovery potential}
\label{sec:discovery}

\begin{figure}[p]
\begin{center}
\includegraphics[scale=0.35,angle=270]{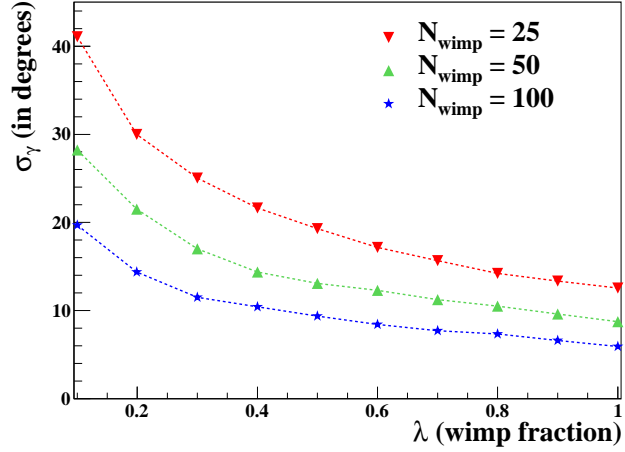}
\begin{center}
\end{center}
 \includegraphics[scale=0.35,angle=270]{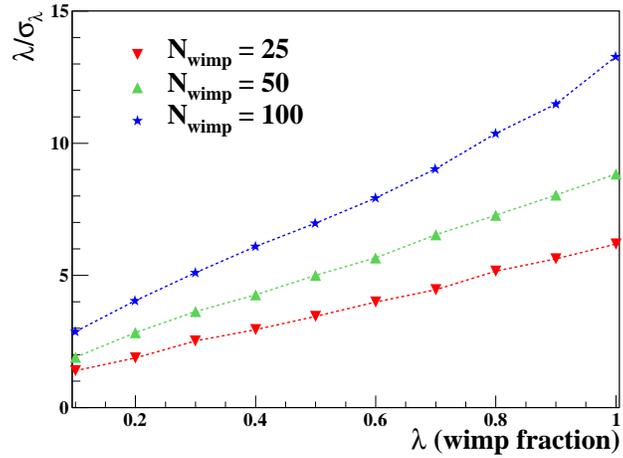}
\caption{Upper panel presents the directional signature $\sigma_{\gamma}$ (in degrees), the radius of the  $68 \ \%$ CL contour of the 
$\mathscr{L}(\ell,b)$ distribution, as a function of the WIMP fraction. 
Lower panel presents the significance ($\lambda/\sigma_\lambda$) as a function of the WIMP fraction $\lambda=S/(S+B)$. 
Results are produced for a $^{19}$F target, a  100 GeV.c$^{-2}$ WIMP 
 and considering recoil energies in the range  5 keV $\leq E_R \leq$ 50 keV.}  
\label{fig:exposition}
\end{center}
 \end{figure}

In order to explore the robustness of the method, a systematic study has been
done with $10^4$ experiments for various number of WIMP events ($N_{\rm wimp}$) 
and several values of WIMP fraction in the observed map ($\rm \lambda$), ranging from 0.1 to 1. 
For a given cross-section, these two parameters are 
related respectively with the exposure and the rejection power of the offline analysis 
preceding  the likelihood method.\\
Figure~\ref{fig:exposition} presents on the upper panel the directional signature, taken as the value of $\sigma_{\gamma} = \sqrt{  \sigma_\ell
\sigma_b }$,  the radius of the  $68 \%$ CL contour of the marginalised
$\mathscr{L}(\ell,b)$ distribution, as a function of $\lambda$. It is related to the ability to 
recover the main signal direction and to sign its Galactic origin. It can first be  noticed that the directional  
signature is of the order of 10$^\circ$ to 20$^\circ$ on a wide range of WIMP fractions.
Even for low number of WIMPs and for a low WIMP fraction (meaning a poor rejection power), 
the directional signature remains clear. From this, we conclude
that a directional evidence in favor of Galactic Dark Matter may be obtained with 
upcoming experiments even at low exposure (i.e. a low number of observed WIMPs) and with a non-negligible background 
contamination (even greater than 50\%).\\
However, a convincing proof of the detection of WIMPs would require  a directional  
signature with  sufficient significance. We defined the significance of this identification strategy as $\lambda/\sigma_\lambda$, presented
on figure~\ref{fig:exposition} (lower panel) as a function of $\lambda=S/(S+B)$. As expected, the significance is increasing both with the 
number of WIMP events and with the WIMP fraction, but we can notice that an evidence ($\rm 3 \sigma$) or a discovery ($\rm 5
\sigma$) of a Dark Matter signal would require either a larger number of WIMPs or a lower background contamination.\\
Using this map-based likelihood method, a directional detector may provide, as a first step, 
a Galactic signature  even with a low number of WIMPs. For instance, a signal pointing towards Cygnus within 
$20^\circ$ can be obtained with as low as 25 WIMPs with a 
50\% background contamination. For an axial cross-section on nucleon of  $\rm \sigma_{n} = 1.5 \times 10^{-3} \ pb$, this 
corresponds to an exposure of 400 kg.day in CF$_4$. In a second step, with an exposure four times larger, 
the directional signature would be only slightly better (10$^\circ$) but the significance would be 
much higher ($\sim 7\sigma$) and the detection much more convincing.

\section{Conclusion}
\label{sec:conclusion}
We have presented a powerful statistical analysis tool to 
extract information from  a  data sample of a directional detector in order to identify a Galactic WIMP signal. 
As a proof of principle, it has been tested within the framework of an isothermal spherical halo model.
We have shown the feasibility to extract from an observed map the main incoming direction of the signal and its significance, thus  proving  its Galactic origin.  
Evidence in favor of Galactic Dark Matter may be within reach of upcoming directional detectors even at low exposure. In a second step, with 
increasing exposure and by using an extended method currently under development, Dark Matter properties (triaxility, halo rotation, WIMP mass) may also be constrained.\\

For this first study of directional detection as a strategy to discover Galactic Dark Matter, the standard halo model has 
been considered.  
It may be noticed that   recent results from high resolution numerical simulations seem to be in favour of non-smooth WIMP velocity
    distributions \cite{vogelsberger,kuhlen}.
    However, as their resolution is many orders of magnitude larger than the scale of the ultra-local Dark Matter distribution probed by current
     and future detectors, this result is not fully relevant for direct detection. Even if applied at ultra-local scale, 
     the result from \cite{kuhlen} is that high velocity WIMPs ($v > 500$~km.s$^{-1}$) could deviate from the Solar motion direction
      ($\ell_\odot ,  b_\odot $) within about 10$^{\circ}$ due to the presence of subhalos, tidal streams or an anisotropic velocity distribution. But, 
      as the minimal speed is 130~km.s$^{-1}$ in this study and as the Maxwellian distribution of WIMP velocity is peaked around 300~km.s$^{-1}$ the contribution
       of high speed
      WIMPs to the WIMP flux is negligible. For instance, this high speed feature would be present in data from detector with a higher energy threshold.
      In such case, this 
      will mildly affect the result presented in this Letter by shifting the main recoil direction recovered from the likelihood analysis by a few degrees.
      However, the discovery proof will  still be  reached as long as the WIMP signal is still directional, 
      thus  remaining  
       different from the background.  Effect of non-standard halo models (triaxial, with stochastic features or streams) will be addressed in a dedicated forthcoming paper.

\section*{Note added after submission}
While this work was being refereed, a related work from A.~M.~Green and 
B.~Morgan \cite{Green:2010zm} appeared on the arXiv. Their paper is complementary to this one as it is estimating the
number of WIMP events required to confirm the direction of Solar motion ($\ell_\odot ,  b_\odot $) as the median inverse recoil direction 
at 95\% CL. 
Then, both this work and the one from \cite{Green:2010zm} highlight the fact that directional detection is a 
key issue to clearly identify a positive Dark Matter detection.

%
%

\end{document}